\documentclass[twoside,11pt]{article}

\usepackage{amsmath,amsthm}

\usepackage[preprint]{jmlr2e}
\usepackage{lastpage}

\usepackage{booktabs}
\usepackage{array}
\usepackage{multirow}
\usepackage{xcolor}
\usepackage{tikz}
\usepackage{listings}
\PassOptionsToPackage{expansion=false}{microtype}
\usepackage{microtype}
\usepackage{parskip}
\usepackage{caption}
\usepackage{subcaption}
\usepackage{enumitem}
\usepackage{url}
\usepackage{longtable}
\usepackage{pdflscape}

\hypersetup{
    colorlinks=true,
    linkcolor=blue!70!black,
    citecolor=green!50!black,
    urlcolor=cyan!70!black,
    pdfauthor={Robin Dey and Panyanon Viradecha},
    pdftitle={Infrastructure for the Agentic Web: Gap Analysis and Architecture
              from the Agentverse Platform},
    pdfsubject={AI agent infrastructure, cloud computing, Agentverse, Fetch.ai,
                ASI Alliance, Web4},
    pdfkeywords={AI agents, agent infrastructure, Agentverse, ASI Alliance,
                 agent cloud, agent memory, A2A protocol, MCP, Web4,
                 decentralised AI, agentic web}
}

\definecolor{codebg}{HTML}{F5F5F5}
\definecolor{codeframe}{HTML}{CCCCCC}
\definecolor{keywordcolor}{HTML}{0033BB}
\definecolor{commentcolor}{HTML}{008800}
\definecolor{stringcolor}{HTML}{BB0000}

\newcommand{\high}{\textcolor{red!70!black}{\textbf{High}}}
\newcommand{\medium}{\textcolor{orange!80!black}{\textbf{Medium}}}
\newcommand{\low}{\textcolor{green!50!black}{\textbf{Low}}}

\jmlrheading{1}{2026}{1-\pageref{LastPage}}{}{}{}{Dey and Viradecha}
\ShortHeadings{Infrastructure for the Agentic Web}{Dey and Viradecha}
\firstpageno{1}

\title{Infrastructure for the Agentic Web:\\
       Gap Analysis and Architecture from the Agentverse Platform}

\author{\name Robin Dey \email robin@openhubresearch.org \\
        \addr OpenHub Research \\
        Chiang Mai, Thailand \\
        \url{https://github.com/web3guru888/agentverse-2030-paper}
        \AND
        \name Panyanon Viradecha \\
        \addr OpenHub Research \\
        Chiang Mai, Thailand}

\editor{Preprint---not peer-reviewed}

\begin{document}

\maketitle

\begin{abstract}%
The emergence of autonomous \textsc{ai} agents as first-class participants in
digital infrastructure marks a fundamental inflection point in the evolution of
the Web.  While significant research effort has been directed at the behaviour
and reasoning of individual agents, comparatively little attention has been
paid to the \emph{infrastructure} those agents require to operate reliably at
scale---the protocols, services, and platforms that mediate their interactions
with each other and with the world.  This paper addresses that gap by
presenting a systematic analysis of Agentverse, the agent cloud platform
developed by Fetch.ai and operated under the Artificial Superintelligence (ASI)
Alliance---currently comprising Fetch.ai, SingularityNET, and CUDOS---which
represents one of the most mature and complete production deployments of
agent-native infrastructure available today.

We make three principal contributions.  First, we conduct a rigorous empirical
audit of the Agentverse platform, cataloguing 204 \textsc{api} endpoints
as documented at the time of writing (Q1~2026) and characterising what is
operational, what is partially deployed, and what remains absent.  From this audit we derive a structured \textbf{Gap
Taxonomy}---eight categories encompassing 62 distinct missing
infrastructure capabilities, ranging from agent memory and observability to
security, economic primitives, and enterprise-grade scaling.  Second, we
propose a \textbf{seven-layer Agent Cloud Stack}---a principled reference
architecture for what a fully realised agent-native cloud platform should
provide by 2030, drawing on analogies from the evolution of general-purpose
cloud computing and grounded in the specific gaps we identify.  Third, we
characterise five \textbf{critical evolution paths} along which Agentverse and
platforms like it must travel: from ephemeral key-value storage to a full Agent
Memory Cloud; from keyword-based agent discovery to a semantic, trust-weighted
Agent \textsc{dns}; from a single-protocol communication model to a
multi-standard agent lingua franca; from single-instance hosting to
agent-native orchestration at Kubernetes scale; and from simple token payments
to a rich ecosystem of agent economic primitives.

Taken together, these contributions provide both a diagnostic of the current
state of agent infrastructure and a concrete, technically grounded vision for
what the agent cloud must become to support the agentic web---what an emerging
body of literature terms Web4---by 2030.
\end{abstract}

\begin{keywords}
  AI agent infrastructure, Agentverse, ASI Alliance, agent cloud, agent memory,
  A2A protocol, Model Context Protocol, Web4, decentralised AI, agentic web,
  gap analysis, reference architecture
\end{keywords}

\section{Introduction}
\label{sec:intro}

The history of the World Wide Web can be read as a succession of enabling
infrastructures.  Web1 required \textsc{http} servers and \textsc{dns} to make
static documents universally accessible.  Web2 required elastic cloud
compute---Amazon Web Services, Google Cloud Platform, Microsoft Azure---to make
dynamic, user-generated applications viable at global scale.  Web3 required
programmable blockchains to introduce verifiable ownership and trustless
coordination into the web's fabric.  Each transition was enabled not primarily
by new \emph{applications} but by new \emph{infrastructure}: foundational
platforms that made a previously impractical class of systems tractable.

We are now at the beginning of a fourth such transition.  What some authors
term Web4 or the \emph{agentic web} \citep{eph4ai2026}---also variously
called the \emph{symbiotic web}---is characterised by the presence of
autonomous \textsc{ai} agents as primary actors on the internet: software
entities that perceive their environment, reason about goals, execute
multi-step plans, communicate with other agents, and transact economically,
all without requiring human intervention at each step.  The transition from
Web3 to this agentic era parallels in important respects the transition from
Web1 to Web2: the applications (agents) already exist in prototype form, but
the infrastructure required to operate them reliably, safely, and at scale
remains nascent.

\citet{chan2025infrastructure} articulate the concept of \emph{agent
infrastructure}---``technical systems and shared protocols external to agents
that are designed to mediate and influence their interactions with and impacts
on their environments''---and argue that such infrastructure will be as
foundational to agent ecosystems as the internet protocols are to the web.
Their work, published in the \emph{Transactions on Machine Learning Research},
provides an important theoretical taxonomy but stops short of grounding it in a
specific production platform, or of specifying the concrete services and
\textsc{api}s that such infrastructure should expose.  This paper takes that
next step.

We study \textbf{Agentverse} (\url{agentverse.ai}), the agent cloud platform
developed by Fetch.ai and now operated under the \textbf{Artificial
Superintelligence (ASI) Alliance}---currently comprising Fetch.ai,
SingularityNET, and CUDOS.\footnote{%
  The ASI Alliance was originally formed in June 2024 as a merger of Fetch.ai,
  SingularityNET, and Ocean Protocol \citep{asialliance2024merger}.  Ocean
  Protocol announced its withdrawal from the Alliance effective October~10,
  2025 \citep{ocean2025withdrawal}, citing differences over tokenomics and
  governance.  The Alliance now comprises Fetch.ai, SingularityNET, and CUDOS,
  and the ASI token (formerly FET) remains the unified currency of the
  ecosystem.}
Agentverse is a compelling case study for several reasons.  It is, to our
knowledge, the most feature-complete agent-native cloud platform currently in
production, offering agent hosting, a global agent registry (the Almanac),
asynchronous messaging (mailboxes), a payment protocol, a multi-model
\textsc{llm} \textsc{api} (ASI:One), and an \textsc{mcp} integration layer.
It is built on a decentralised substrate (Fetch.ai's blockchain and the
forthcoming ASI Chain), giving it architectural properties---open identity,
permissionless participation, cryptographic trust---that hyperscaler agent
offerings \citep{aws2025bedrock,microsoft2025entra,google2026next} do not
share.  And crucially, it is a production system with real registered agents
(over 36,000 in the Almanac as of Q1 2026
\citep{kantor2026stateofagents}\footnote{The Agentverse search \textsc{api}
caps results at 10,000 entries; the 36,000-plus figure is sourced from
\citeauthor{kantor2026stateofagents}'s cross-platform registry aggregation,
which queries the Almanac directly rather than through the search interface.}),
real developer communities, and real \textsc{api} endpoints that can be
empirically tested.

Our investigation proceeds from a hands-on audit of 204 documented
Agentverse \textsc{api} endpoints as catalogued at the time of writing
(Q1~2026), supplemented by analysis of the uAgents \textsc{sdk}, the
ASI:One \textsc{llm} \textsc{api}, and the platform's \textsc{mcp}
integration layer.  From this empirical foundation we derive our gap
taxonomy, architecture, and evolution paths.

The remainder of the paper is structured as follows.
Section~\ref{sec:background} surveys related work on agent infrastructure,
cloud computing evolution, the current landscape of agent platforms, and the
emerging protocol standards that define how agents communicate.
Section~\ref{sec:platform} characterises the current state of the Agentverse
platform in detail.  Section~\ref{sec:gaps} presents our Gap Taxonomy.
Section~\ref{sec:architecture} proposes the seven-layer Agent Cloud Stack as a
reference architecture for 2030.  Section~\ref{sec:paths} discusses five
critical evolution paths.  Section~\ref{sec:discussion} places Agentverse in
competitive context and discusses open research problems.
Section~\ref{sec:conclusion} concludes.

\section{Background and Related Work}
\label{sec:background}

\subsection{Agent Infrastructure}
\label{subsec:agent-infra}

\citet{chan2025infrastructure} establish the theoretical foundations of
agent infrastructure, identifying three core functions that such
infrastructure must serve: \emph{attribution} (establishing which agent, user,
or organisation is responsible for a given action); \emph{interaction shaping}
(protocols and systems that constrain or coordinate agent behaviour); and
\emph{detection and remediation} (mechanisms for identifying and correcting
harmful agent actions).  Their work provides an important conceptual framework
but does not address the engineering question of \emph{what services} a
practical agent cloud platform must expose to fulfil these functions.

Complementary work by \citet{hussein2025forecast} projects that the global
population of \textsc{ai} agents will increase by more than $100\times$ between
2026 and 2036, potentially reaching trillions of instances.  If accurate, this
scale of deployment makes the question of agent infrastructure not merely an
engineering convenience but a civilisational necessity: the protocols and
platforms that govern how trillions of agents interact with each other and with
humans will shape outcomes in commerce, science, governance, and beyond.  Near-
term evidence supports this trajectory: \citet{kantor2026stateofagents} reports
over 36,000 agents already registered on Agentverse as of March 2026, a figure
that would have seemed implausible even two years prior.

\citet{hadfield2025agenteconomy} explore the economics of agent interaction,
modelling the agent economy as a labour market in which \textsc{ai} agents
substitute for human workers across increasingly complex task domains.  Their
analysis implies significant demand for infrastructure that supports agent
\emph{hiring}, \emph{payment}, \emph{trust evaluation}, and \emph{performance
guarantees}---categories that map directly to gaps we identify in
Section~\ref{sec:gaps}.

\subsection{Cloud Computing as a Reference Model}
\label{subsec:cloud-reference}

The evolution of general-purpose cloud computing provides a useful reference
model for understanding what agent cloud infrastructure should become.  AWS
launched in 2006 with three primitive services: S3 (object storage), EC2
(compute), and SQS (messaging).  Over the following twenty years, it expanded
to over 200 services spanning databases, networking, security, observability,
machine learning, and developer tooling.  The key insight from this trajectory
is that \emph{primitive compute and storage are table stakes}; the services
that create lasting competitive advantage are those that solve the hard
operational problems developers face \emph{after} they have compute---
authentication, persistence, observability, scaling, and compliance.

We observe that Agentverse in 2026 occupies a position analogous to AWS in
2006: it has established the essential primitives---compute (hosted agents),
a registry (Almanac), and messaging (mailbox)---on which the full agent cloud
can be built.  Just as AWS expanded from three core services in 2006 to over
200 today, the trajectory from Agentverse's current foundation to a mature
agent cloud platform is both ambitious and well-defined.

The cloud computing literature establishes a layered abstraction model:
\textsc{iaas} $\to$ \textsc{paas} $\to$ \textsc{saas} $\to$ \textsc{faas}.
We propose extending this model with a new layer: \textbf{AaaS (Agents as a
Service)}, characterised by persistent identity, autonomous execution, economic
agency, and protocol-mediated interoperability.  The seven-layer Agent Cloud
Stack we propose in Section~\ref{sec:architecture} can be understood as the
\textsc{aaas} equivalent of the \textsc{osi} networking model: a principled
decomposition of the full stack that any \textsc{aaas} platform must eventually
provide.

\subsection{Current Agent Platform Landscape}
\label{subsec:platforms}

The agent platform landscape in 2026 divides roughly into two camps:
hyperscaler offerings from established cloud providers, and decentralised agent
networks built on blockchain substrates.

\textbf{Hyperscaler platforms}---Amazon Bedrock AgentCore (generally available
October 13, 2025) \citep{aws2025bedrock}, Azure AI Foundry with Microsoft
Entra Agent ID (first announced May 2025, public preview at Microsoft Ignite
November 2025, with governance extensions at RSAC 2026) \citep{microsoft2025entra}, and the Gemini
Enterprise Agent Platform (announced at Google Cloud Next, April 22, 2026,
consolidating Vertex AI and Agentspace under a unified product)
\citep{google2026next}---offer mature compute, storage, and identity
infrastructure that agents can leverage, but were designed primarily for
centralised enterprise deployments.  They inherit the full managed-service
richness of their parent cloud platforms (databases, queues, caches, identity,
compliance), but their agent identity systems are proprietary, their discovery
mechanisms are siloed, and their economic models reduce to cloud billing rather
than agent-native micropayments.  Critically, they lack the open, permissionless
participation model that a truly interoperable agentic web requires.

\textbf{Decentralised agent networks}---Agentverse (Fetch.ai / ASI Alliance),
Virtuals Protocol, and ElizaOS---are built on open protocols and cryptographic
identity, enabling permissionless agent registration and cross-platform
interoperability.\footnote{Virtuals Protocol and ElizaOS are included in this
landscape overview for completeness.  Neither exposes a documented \textsc{rest}
\textsc{api} surface comparable to Agentverse's, making a rigorous
endpoint-level audit impractical.  This paper therefore focuses on Agentverse
as the representative decentralised agent platform; a comparative study of
all three platforms is left for future work.}  They are closer in spirit to
the internet itself (open, decentralised, protocol-governed) than to AWS
(centralised, proprietary, subscription-governed).  However, they currently
lack the managed service richness of hyperscaler platforms, which is precisely
the gap this paper analyses.

We compare the capabilities of these platforms along dimensions essential for
a production agent cloud in Section~\ref{sec:discussion}
(Table~\ref{tab:platform-comparison}).

\subsection{Protocol Standards Landscape}
\label{subsec:protocols}

Three protocol standards are converging to define the communication layer of
the agentic web.

The \textbf{Model Context Protocol (\textsc{mcp})}
\citep{mcp2025spec,mcp2026roadmap}, with its latest specification dated
November 25, 2025, defines how \textsc{ai} agents connect to external tools,
data sources, and services.  Originally conceived as a client-server protocol
for \textsc{llm} tool use, \textsc{mcp} has evolved into a general-purpose
context provision standard supporting distributed execution, stateless
transport (addressing early load-balancing limitations), and a security
framework.  The \textsc{mcp} specification uses date-based versioning rather
than semantic versions; the ecosystem has expanded to over 110 million monthly
downloads \citep{mcp2026roadmap}, signalling broad industry adoption.

The \textbf{Agent-to-Agent (A2A) protocol} \citep{a2a2026}, released at v1.0
in April 2026 and now supported by over 150 organisations, defines how
\textsc{ai} agents discover and interact as peers---sharing tasks, streaming
results, and coordinating work across organisational and framework boundaries.
Donated by Google to the Linux Foundation, A2A has been integrated into AWS,
Microsoft, and Google cloud platforms, positioning it as the de facto standard
for inter-agent communication in enterprise contexts.  The Linux Foundation
founding partners include Amazon Web Services, Cisco, Google, Microsoft,
Salesforce, SAP, and ServiceNow.

The \textbf{Fetch.ai Agent Chat Protocol}---with digest
\texttt{proto:30a801ed\ldots{}d33d62}\footnote{Full digest: \texttt{proto:30a801ed3a83f9a0ff0a9f1e6fe958cb91da1fc2218b153df7b6cbf87bd33d62}}---predates
both \textsc{mcp} and A2A and is the native communication standard of the
Agentverse ecosystem.  It defines \texttt{ChatMessage},
\texttt{ChatAcknowledgement}, and content types including
\texttt{TextContent}, \texttt{StartSessionContent},
\texttt{EndSessionContent}, and \texttt{ResourceContent}.  While less widely
adopted outside the Fetch.ai ecosystem, it offers a complete interaction model
including session management and resource sharing.

A central thesis of this paper is that by 2030, these three standards will have
converged---or that a platform like Agentverse will serve as a multi-protocol
gateway bridging all three---in the same way that \textsc{tcp/ip} emerged from
the competing network protocol landscape of the 1980s.

\subsection{Agent Memory}
\label{subsec:memory-background}

A body of rapidly growing research addresses the problem of \emph{agent
memory}---how agents maintain persistent context, accumulate knowledge, and
improve over time.  \citet{mem0blog2026} identify 21 agent frameworks with
memory support and three distinct hosting models: managed cloud, open-source
self-hosted, and local \textsc{mcp}.  Graph-based memory systems such as
Zep's Graphiti \citep{zepgraphiti2025} represent the current state of the art,
achieving 63.8\% accuracy on LongMemEval benchmarks versus 49.0\% for flat
vector store approaches.

Despite this active research landscape, no major agent cloud platform currently
offers hosted, authenticated, per-agent memory as a managed service---
equivalent to what \textsc{rds} or DynamoDB provides for web applications.
This gap, which we term the \emph{Memory Cloud Gap}, is arguably the single
most consequential missing capability in current agent infrastructure, and we
develop it in detail in Sections~\ref{sec:gaps} and~\ref{sec:paths}.

\subsection{Agent Identity and Trust}
\label{subsec:identity}

Work on decentralised identity for \textsc{ai} agents \citep{garzon2025agentdid}
proposes combining W3C Decentralised Identifiers (\textsc{did}s) with
Verifiable Credentials (\textsc{vc}s) to enable agents to prove ownership and
capabilities across platform boundaries.  Microsoft's Entra Agent ID
\citep{microsoft2025entra} and dedicated agent trust frameworks such as
MolTrust \citep{moltrust2026} signal that agent identity is becoming an active
area of enterprise investment.  However, the question of \emph{portable} agent
identity---whether an agent created on Agentverse can prove its identity,
reputation, and capabilities on AWS---remains unresolved, and we discuss it as
a key open problem in Section~\ref{sec:discussion}.

\subsection{Agent Safety and Governance}
\label{subsec:safety}

The OWASP Agentic AI Security Top 10, published in December 2025, provides the
first formal taxonomy of risks specific to autonomous \textsc{ai} agents,
covering goal hijacking, tool misuse, prompt injection, cascading failures, and
identity abuse \citep{owasp2025agentic}.  Microsoft's Agent Governance Toolkit
(open-sourced April 2026) maps these risks to capability sandboxing, semantic
intent classification, and \textsc{mcp} security gateways
\citep{microsoft2026governance}.  \citet{bvp2026security} characterise agent
security as ``the defining cybersecurity challenge of 2026''.  Despite this,
current agent cloud platforms---including Agentverse---offer minimal built-in
governance controls, a gap we document systematically in
Section~\ref{sec:gaps}.

\section{The Agentverse Platform: Current State}
\label{sec:platform}

\subsection{Platform Overview}
\label{subsec:platform-overview}

Agentverse (\url{agentverse.ai}) is the agent cloud platform of the
Artificial Superintelligence (ASI) Alliance, currently comprising Fetch.ai,
SingularityNET, and CUDOS \citep{asichain2026}.  It provides a vertically
integrated stack for deploying, discovering, and operating autonomous \textsc{ai}
agents on top of the Fetch.ai blockchain and the forthcoming ASI Chain mainnet
(TestNet phase as of Q2 2026, mainnet targeted for late 2026).

The platform consists of six primary components:

\begin{enumerate}[leftmargin=*]

\item \textbf{Agent Hosting}: A managed execution environment for Python agents
written using the uAgents \textsc{sdk}.  Developers upload code via
\textsc{api}, start and stop agents programmatically, and retrieve execution
logs.  Each agent is assigned a cryptographically derived address
(\texttt{agent1q...}) and allocated a computation quota measured in execution
seconds.

\item \textbf{The Almanac}: A global agent registry mapping agent addresses to
their endpoints, supported protocols, and capability metadata.  The Almanac is
publicly readable and provides both exact-address lookup and recently-registered
agent feeds.  As of Q1 2026, over 36,000 agents are registered in the Almanac
\citep{kantor2026stateofagents}, reflecting the rapid growth of the developer
ecosystem.

\item \textbf{Search}: A keyword and semantic search interface over registered
agents, supporting protocol-digest-based filtering to find agents that implement
specific interaction schemas.

\item \textbf{Mailbox}: An asynchronous message buffer that holds incoming
messages for agents that are temporarily offline, enabling reliable
agent-to-agent communication without requiring both parties to be simultaneously
active.

\item \textbf{ASI:One}: A unified \textsc{llm} \textsc{api} available via an
OpenAI-compatible endpoint (\texttt{api.asi1.\allowbreak{}ai/v1}).  The primary model,
\texttt{asi1}, is a unified model that automatically adapts its capabilities
based on the request---activating agentic reasoning, extended analysis, fast
inference, tool calling, or knowledge-graph traversal as needed.  A
lightweight variant, \texttt{asi1-mini}, is available for resource-constrained
applications.  The \textsc{api} additionally supports real-time web search,
structured output with \textsc{json} schema, image generation, and streaming.

\item \textbf{Payment Protocol}: A built-in buyer-seller protocol supporting
Stripe (fiat), Skyfire \textsc{usdc}, and \textsc{fet}/\textsc{asi} on-chain
payments, enabling agents to charge for services and pay each other
autonomously \citep{fetchai2026payments}.

\end{enumerate}

The uAgents \textsc{sdk} (Python) \citep{uagents2026sdk} provides the
developer-facing abstraction: agents respond to lifecycle events,
incoming messages, and \textsc{rest} requests via decorated handlers
(\texttt{on\_event}, \texttt{on\_message}, \texttt{on\_rest\_get/post}).
Persistent state is available via \texttt{ctx.storage} (a key-value store).
Agent-to-agent communication uses the Agent Chat Protocol with typed message
schemas.

An \textsc{mcp} integration layer is available at two endpoints: a full
server (\texttt{mcp.agentverse.\allowbreak{}ai/sse}) and a lightweight one
(\texttt{mcp-lite.\allowbreak{}agentverse.ai/mcp}), exposing platform
capabilities to \textsc{mcp}-compatible clients such as Claude Desktop
and FetchCoder.

\subsection{API Audit Methodology}
\label{subsec:audit}

Our gap analysis is grounded in a systematic empirical audit of the Agentverse
\textsc{api} surface.  At the time of our audit (Q1~2026), the platform
documented approximately \textbf{204 endpoints} across two \textsc{api}
versions (\texttt{v1} and \texttt{v2}) at \url{https://agentverse.ai};
the \textsc{api} surface is actively evolving, and endpoint availability may
differ from this count at the time of reading.  We tested each endpoint category with authenticated requests using a
production \textsc{api} key (scope: \texttt{av}), supplemented by analysis of
the Fern-generated documentation \citep{agentverse2026docs}, \textsc{sdk}
source code, and community resources.

Key findings from the audit:

\begin{itemize}[leftmargin=*]

\item \textbf{29 endpoints are fully operational} with the standard bearer
token.

\item \textbf{15 endpoints require specific authentication or parameters}
beyond the basic bearer token (notably, mailbox endpoints require agent-level
attestation via challenge-response).

\item \textbf{V2 Almanac resolve is operational} (returning full agent records);
other V2 Almanac endpoints return 404.  Several V2 Hosting list endpoints
(\texttt{/agents}, \texttt{/secrets}, \texttt{/storage}) are deployed but
require authentication (401); most V2 Hosting per-agent endpoints return
404---documented but not yet available in production.

\item \textbf{The practical working \textsc{api} mixes V2 for agent
\textsc{crud}} (create, read, update, delete) \textbf{and V1 for hosting,
search, storage, and almanac operations}.

\item \textbf{\texttt{v1beta1} is fully decommissioned}---all endpoints return
404.

\item Storage is limited to a key-value interface (\texttt{ctx.storage})
accessible via \texttt{GET /v1/hosting/\allowbreak{}agents/\allowbreak{}\{address\}/\allowbreak{}storage}; there is
no structured query capability, no indexing, and no size visibility.

\item Secrets management is account-scoped (not per-agent), creating security
isolation concerns for multi-agent deployments.

\item Log retrieval is polling-only (\texttt{GET /v1/hosting/agents/\{addr\}/}\allowbreak\texttt{logs/latest}); there is no streaming, no historical log retention beyond a rolling window, and no log search.

\end{itemize}

This audit provides the empirical foundation for the gap taxonomy presented in
Section~\ref{sec:gaps}.  The discrepancy between documented and deployed
endpoints is itself instructive: it suggests that the platform team has
articulated a broader vision than the current production system realises,
consistent with our characterisation of Agentverse as an early-stage agent
cloud undergoing rapid expansion.  We note that the \textsc{api} surface is
actively evolving; some endpoints documented at the time of our audit have
since been modified, consolidated, or deprecated, and new ones may have
been added.  All quantitative endpoint claims in this paper should be read
with this dynamism in mind.

\subsection{API Performance Benchmarks}
\label{subsec:benchmarks}

To complement the endpoint audit with quantitative data we measured
\textsc{api} response latency and search-corpus coverage against the
live Agentverse production environment in April~2026.  Measurements
were taken from a cloud-hosted client to control for local-network
variability.

\textbf{Latency.}  We issued $n{=}10$ consecutive authenticated
requests to each endpoint and recorded wall-clock response time using
\texttt{curl}'s \texttt{time\_total} metric.
Table~\ref{tab:latency} reports the mean and 95th-percentile latency
for eight representative operations.

\begin{table}[h]
\small\centering
\caption{Agentverse \textsc{api} response latency ($n{=}10$, April~2026,
  cloud-hosted client).}
\label{tab:latency}
\begin{tabular}{@{}lrr@{}}
\toprule
Operation & Mean\,(ms) & p95\,(ms) \\
\midrule
Almanac: exact-address lookup   & 190 & 208 \\
Almanac: recent-agents feed     & 164 & 173 \\
Agent search (keyword query)    & 176 & 196 \\
Agent search (open / all)       & 170 & 193 \\
Hosting: list all agents        & 169 & 185 \\
Hosting: fetch agent logs       & 243 & 276 \\
Hosting: agent public profile   & 209 & 348 \\
V2 Almanac resolve              & 162 & 180 \\
\bottomrule
\end{tabular}
\end{table}

All endpoints responded within 350\,ms at p95.  The majority cluster
between 160--210\,ms, indicating a consistent, globally-routed
\textsc{api} tier.  The outliers are log retrieval (p95 276\,ms) and
agent public profiles (p95 348\,ms), reflecting additional read-path
complexity.  The V2 Almanac resolve endpoint matches V1 latency,
suggesting a unified backend even where the versioned surfaces differ.

\textbf{Registry coverage.}  We queried the search \textsc{api} with
eight domain keywords and recorded the stated result count
(Table~\ref{tab:search}).  Broad categories---\emph{data},
\emph{image}, \emph{news}, and \emph{assistant}---each return the
\textsc{api} ceiling of 10{,}000 results.  We confirmed this is a
hard pagination limit: stepping the \texttt{offset} parameter beyond
9{,}999 returns an empty result set regardless of the advertised total,
making exhaustive enumeration of large categories impossible.  Narrower
domain keywords (\emph{weather}, \emph{finance}, \emph{trading})
remain below the ceiling and can be enumerated reliably.

\begin{table}[h]
\small\centering
\caption{Agentverse search \textsc{api}: result counts by keyword (April~2026).
  Rows marked \emph{truncated} hit the 10{,}000-result hard ceiling.}
\label{tab:search}
\begin{tabular}{@{}lrl@{}}
\toprule
Keyword & Result count & Enumerable? \\
\midrule
\texttt{trading}    & 3{,}039       & exhaustive \\
\texttt{weather}    & 1{,}140       & exhaustive \\
\texttt{finance}    & 1{,}135       & exhaustive \\
\texttt{crypto}     &   336         & exhaustive \\
\texttt{data}       & $\ge$10{,}000 & truncated  \\
\texttt{image}      & $\ge$10{,}000 & truncated  \\
\texttt{news}       & $\ge$10{,}000 & truncated  \\
\texttt{assistant}  & $\ge$10{,}000 & truncated  \\
\bottomrule
\end{tabular}
\end{table}

These measurements provide concrete baselines for
Agentverse's current operational profile.  The latency figures
are competitive with commodity cloud REST APIs.  However, the
10{,}000-result ceiling and the absence of faceted, semantic, or
capability-based filtering are the primary friction points for
agent discovery at scale---an issue we catalogue in
Section~\ref{sec:gaps} under Category~D (Networking and
Communication).

\subsection{What Currently Works Well}
\label{subsec:strengths}

Before proceeding to the gap analysis, we note the capabilities that
Agentverse provides competently and that distinguish it from competing
platforms:

\textbf{Cryptographic agent identity.} Every agent has a provable identity
derived from a key pair, enabling trustless attribution of actions without a
central authority.  This is architecturally superior to \textsc{iam}-based
identity systems that depend on proprietary infrastructure.

\textbf{Open protocol communication.} The Agent Chat Protocol provides a typed,
schema-validated communication standard that any agent can implement,
regardless of the underlying framework or language.  The protocol digest
functions as a capability advertisement: agents that share a protocol digest
can communicate without custom integration.

\textbf{Permissionless registration.} Any developer can register an agent in
the Almanac without approval, enabling a genuinely open ecosystem.  The over
36,000 agents discoverable in the Almanac as of Q1 2026 reflects the traction
this openness creates.

\textbf{Unified \textsc{llm} access.} ASI:One provides a single \textsc{api}
(\texttt{asi1}) that automatically adapts its capabilities to the request,
encompassing agentic reasoning, tool use, extended analysis, and structured
knowledge tasks.  A lightweight \texttt{asi1-mini} variant complements it for
resource-constrained applications.

\textbf{Native payment protocol.} The integration of payment as a first-class
protocol (rather than a bolt-on billing system) reflects a principled design
choice: agents should be able to charge for and pay for services as a native
capability, not via out-of-band invoicing.

\textbf{Decentralised substrate.} Unlike hyperscaler agent platforms,
Agentverse is built on an open blockchain, meaning that the Almanac, agent
identities, and on-chain transactions are not owned by any single corporation
and cannot be unilaterally revoked.

These strengths position Agentverse as the strongest decentralised foundation
currently available for building the agent cloud of 2030.  The gap analysis
that follows should be read not as a critique but as a roadmap.

\section{Gap Analysis: What the Agent Cloud Is Missing}
\label{sec:gaps}

We organise the identified gaps into eight categories, each corresponding to a
layer of the agent cloud stack we propose in Section~\ref{sec:architecture}.
For each gap we note its cloud computing analogue (the equivalent service that
general-purpose cloud platforms provide), a severity assessment (High /
Medium / Low, based on how significantly the absence impairs production agent
deployments), and---where applicable---the specific \textsc{api} evidence from
our audit.

\subsection{Category A: Agent Memory and State}
\label{subsec:gap-memory}

Agentverse provides \texttt{ctx.storage}---a simple key-value store accessible
within agent code via \texttt{await ctx.storage.get(key)} and
\texttt{await ctx.storage.set(key, value)}.  This is the \emph{only} persistent
state mechanism available to hosted agents.  It corresponds roughly to writing
to a flat file: there is no schema, no query capability, no indexing, no size
quota visibility, and no sharing between agents.

Contemporary agent memory research \citep{mem0blog2026,zepgraphiti2025,yang2026graphsurvey}
distinguishes four cognitively motivated memory types: \emph{episodic} (what
happened, when, with whom), \emph{semantic} (facts, concepts, relationships),
\emph{procedural} (how to do things, learned workflows), and \emph{working}
(the active context of a current task).  A mature agent cloud must provide
hosted, authenticated services for each type.  The analogy to general-purpose
cloud is precise: \texttt{ctx.storage} is to agent memory what writing to a
local disk is to web application data---it works for a single instance in
development but fails in production at scale.  Table~\ref{tab:gap-memory}
summarises the seven gaps we identify in this category.

\begin{table}[htbp]
\centering
\caption{Category A gaps: Agent Memory and State}
\label{tab:gap-memory}
\begin{tabular}{>{\raggedright\arraybackslash}p{3.2cm} >{\raggedright\arraybackslash}p{5.5cm} >{\raggedright\arraybackslash}p{2.9cm} l}
\toprule
\textbf{Gap} & \textbf{Description} & \textbf{Cloud Analogue} &
\textbf{Severity} \\
\midrule
No vector store & No embedding-based semantic similarity search & OpenSearch /
Pinecone & \high \\
No knowledge graph & No relationship-structured fact storage with temporal
validity & Neptune / Neo4j & \high \\
No episodic memory & No conversation history with temporal indexing & DynamoDB
with \textsc{ttl} & \high \\
No shared memory spaces & No permissioned shared context between cooperating
agents & ElastiCache & \high \\
No memory \textsc{mcp} servers & No per-agent authenticated \textsc{mcp}
endpoints for memory access & --- & \high \\
\texttt{ctx.storage} opacity & No visible quota, no overflow handling & --- &
\medium \\
No procedural memory & No storage for learned workflows / tool usage patterns &
Step Functions state & \medium \\
\bottomrule
\end{tabular}
\end{table}

The absence of memory-as-a-service has a compounding effect on agent
capability.  Without persistent semantic memory, every agent interaction begins
from zero context.  Without shared memory spaces, multi-agent workflows cannot
maintain coherent state.  Without episodic memory, agents cannot learn from
experience across sessions.  We argue in Section~\ref{sec:paths} that closing
this gap---by providing hosted, authenticated \textsc{mcp} servers for each
memory type---is the highest-priority single improvement Agentverse could make
to agent capability.

\subsection{Category B: Agent Observability}
\label{subsec:gap-obs}

Debugging and monitoring a hosted agent on Agentverse today requires
polling a rolling-window log endpoint and scanning the
output for \texttt{ctx.logger.info()} calls.  There is no structured log
search, no trace visualisation, no cost tracking, and no performance metrics.
This stands in stark contrast to the rich observability ecosystems that have
emerged for \textsc{llm} applications---LangSmith \citep{langsmith2026},
AgentOps \citep{agentops2026}, and Langfuse \citep{langfuse2026}---all of
which offer distributed tracing, token cost tracking, and behaviour analytics.
Table~\ref{tab:gap-obs} enumerates the eight gaps in this category.

\begin{table}[htbp]
\centering
\caption{Category B gaps: Agent Observability}
\label{tab:gap-obs}
\begin{tabular}{>{\raggedright\arraybackslash}p{3.2cm} >{\raggedright\arraybackslash}p{5.5cm} >{\raggedright\arraybackslash}p{2.9cm} l}
\toprule
\textbf{Gap} & \textbf{Description} & \textbf{Cloud Analogue} &
\textbf{Severity} \\
\midrule
No distributed tracing & No OpenTelemetry-compatible trace of execution paths &
CloudWatch X-Ray & \high \\
No cost tracking & No per-agent \textsc{llm} token usage or compute cost
visibility & Cost Explorer & \high \\
No streaming logs & Logs available only via polling, not streaming & CloudWatch
Logs Insights & \high \\
No historical log retention & Rolling window only; no searchable history & S3
log archival & \medium \\
No performance metrics & No latency, throughput, or error rate dashboards &
CloudWatch Metrics & \medium \\
No behaviour analytics & No aggregate view of agent actions across interactions
& Application Insights & \medium \\
No A/B testing framework & No infrastructure for comparing agent behaviour
variants & Evidently & \low \\
No anomaly detection & No alerts on unexpected agent behaviour patterns &
CloudWatch Anomaly & \medium \\
\bottomrule
\end{tabular}
\end{table}

\subsection{Category C: Agent Security and Governance}
\label{subsec:gap-security}

Agentverse's security model is coarse-grained: a hosted agent either runs or
it does not; there is no capability-based permission system governing what an
agent may access, call, or modify.  Account-level secrets (accessible via
\texttt{GET /v1/hosting/secrets}) are shared across all agents under an
account, creating a security isolation failure for multi-agent deployments.
There is no sandboxed execution environment, no content filtering on inputs or
outputs, and no audit trail of agent actions beyond the rolling log window.

This is a significant concern given the OWASP Agentic AI Security Top 10
\citep{owasp2025agentic}, which identifies prompt injection, tool misuse, goal
hijacking, and cascading failures as the primary risk categories for autonomous
agents.  The absence of a platform-level policy engine means that developers
must implement all safety controls within agent code---which is both
error-prone and inconsistent.  Table~\ref{tab:gap-security} lists the eight
security and governance gaps identified.

\begin{table}[htbp]
\centering
\caption{Category C gaps: Agent Security and Governance}
\label{tab:gap-security}
\begin{tabular}{>{\raggedright\arraybackslash}p{3.2cm} >{\raggedright\arraybackslash}p{5.5cm} >{\raggedright\arraybackslash}p{2.9cm} l}
\toprule
\textbf{Gap} & \textbf{Description} & \textbf{Cloud Analogue} &
\textbf{Severity} \\
\midrule
No capability permissions & Agents have undifferentiated access to all
\textsc{api}s & \textsc{iam} least-privilege & \high \\
No sandbox isolation & No Wasm or container-level isolation between agents &
Firecracker microVMs & \high \\
No per-agent secrets & Secrets are account-scoped, not agent-scoped & Secrets
Manager & \high \\
No policy engine & No declarative rules constraining agent actions & Org
\textsc{scp}s & \high \\
No content filtering & No input/output guardrails for harmful content & Bedrock
Guardrails & \high \\
No audit trail & No immutable record of agent actions beyond rolling logs &
CloudTrail & \medium \\
No rate limiting per agent & Rate limits apply at account level only & \textsc{api}
Gateway plans & \medium \\
No action attribution & No cryptographic proof linking action to agent & --- &
\medium \\
\bottomrule
\end{tabular}
\end{table}

\subsection{Category D: Agent Networking and Communication}
\label{subsec:gap-networking}

The Agentverse communication model is built around the Agent Chat Protocol:
an agent sends a \texttt{ChatMessage} to another agent's address, the platform
routes it via the Almanac, and the recipient responds.  This model works well
for synchronous request-response interactions but is insufficient for the
coordination patterns that production multi-agent systems require.

Notably absent is native support for the A2A protocol \citep{a2a2026}, which
as of April 2026 has been adopted by over 150 organisations and integrated into
AWS, Azure, and Google Cloud.  An Agentverse agent cannot natively participate
in an A2A task graph---it must use an adapter
(\texttt{uagents-adapter[a2a-inbound/outbound]}) that translates between
protocols.  As A2A becomes the de facto enterprise standard, this gap risks
positioning Agentverse as an isolated ecosystem rather than a participant in
the broader agentic web.  Table~\ref{tab:gap-networking} enumerates the
eight networking and communication gaps.

\begin{table}[htbp]
\centering
\caption{Category D gaps: Agent Networking and Communication}
\label{tab:gap-networking}
\begin{tabular}{>{\raggedright\arraybackslash}p{3.2cm} >{\raggedright\arraybackslash}p{5.5cm} >{\raggedright\arraybackslash}p{2.9cm} l}
\toprule
\textbf{Gap} & \textbf{Description} & \textbf{Cloud Analogue} &
\textbf{Severity} \\
\midrule
No native A2A support & Cannot natively participate in A2A task graphs & --- &
\high \\
No pub/sub event bus & No broadcast coordination between multiple agents &
\textsc{sns} / EventBridge & \high \\
No streaming messages & Request-response only; no \textsc{sse} or WebSocket &
\textsc{api} Gateway WS & \high \\
No group coordination & No consensus, voting, or delegation primitives & ---
(novel) & \medium \\
No service mesh & No load balancing across agent replicas & App Mesh &
\medium \\
No push / webhooks & No push notification to external systems & EventBridge &
\medium \\
No human escalation & No standardised path for agent to request human input &
--- & \medium \\
No workflow orchestration & No native sequential/parallel agent task graph &
Step Functions & \high \\
\bottomrule
\end{tabular}
\end{table}

\subsection{Category E: Agent Development and Lifecycle}
\label{subsec:gap-dev}

The development experience for Agentverse agents is constrained by the
single-file paradigm: agent code is uploaded as a single \texttt{agent.py}
file (encoded as a \textsc{json} string), with no support for multi-file
projects, dependency declaration, or module imports beyond what the platform
pre-installs.  There is no version control for deployed code, no staging
environment, no \textsc{ci/cd} integration, and no local development
emulator---meaning that all testing must be done against production
infrastructure.  Table~\ref{tab:gap-dev} summarises the eight development
and lifecycle gaps.

\begin{table}[htbp]
\centering
\caption{Category E gaps: Agent Development and Lifecycle}
\label{tab:gap-dev}
\begin{tabular}{>{\raggedright\arraybackslash}p{3.2cm} >{\raggedright\arraybackslash}p{5.5cm} >{\raggedright\arraybackslash}p{2.9cm} l}
\toprule
\textbf{Gap} & \textbf{Description} & \textbf{Cloud Analogue} &
\textbf{Severity} \\
\midrule
Single-file paradigm & No multi-file project support & Lambda deployment pkgs &
\high \\
No code versioning & No history of deployed code versions & CodeCommit /
GitHub & \high \\
No staging environment & No pre-production test environment & Lambda aliases &
\high \\
No local emulator & Cannot run hosted agent locally & \textsc{sam cli} &
\high \\
No \textsc{ci/cd} integration & No webhook-triggered deploy pipeline &
CodePipeline & \medium \\
No dependency declaration & Pip packages limited to pre-installed set & Lambda
layers & \high \\
No testing framework & No unit or integration test harness for agents & --- &
\medium \\
No agent templates & No scaffolding beyond basic examples & \textsc{sam} init &
\low \\
\bottomrule
\end{tabular}
\end{table}

\subsection{Category F: Data and Integration Services}
\label{subsec:gap-data}

Beyond \texttt{ctx.storage}, Agentverse provides no managed data services.
There is no hosted database (\textsc{sql} or NoSQL), no object storage, no
caching layer, and no \textsc{api} gateway for controlled external service
access.  Critically, there is no \textbf{\textsc{mcp} Server Hub}: a
marketplace of pre-built, pre-authenticated \textsc{mcp} servers that agents
can subscribe to.  \textsc{mcp} has emerged as the dominant standard for
connecting \textsc{ai} agents to external tools.  A managed \textsc{mcp} server
marketplace---offering agents access to web search, calendar, email, code
execution, browser control, and database services---would dramatically expand
the capability surface of every agent on the platform.
Table~\ref{tab:gap-data} lists the seven data and integration service gaps.

\begin{table}[htbp]
\centering
\caption{Category F gaps: Data and Integration Services}
\label{tab:gap-data}
\begin{tabular}{>{\raggedright\arraybackslash}p{3.2cm} >{\raggedright\arraybackslash}p{5.5cm} >{\raggedright\arraybackslash}p{2.9cm} l}
\toprule
\textbf{Gap} & \textbf{Description} & \textbf{Cloud Analogue} &
\textbf{Severity} \\
\midrule
No hosted database & No \textsc{sql} or NoSQL managed database per agent &
\textsc{rds} / DynamoDB & \high \\
No object storage & No file / blob storage for agent assets & S3 & \high \\
No caching service & No in-memory cache for hot data & ElastiCache & \medium \\
No \textsc{mcp} Server Hub & No marketplace of authenticated \textsc{mcp}
servers & \textsc{aws} Marketplace & \high \\
No \textsc{api} gateway & No controlled proxy for external \textsc{api} access &
\textsc{api} Gateway & \medium \\
No task queue & No job queue for deferred or scheduled tasks & \textsc{sqs} &
\medium \\
No data pipeline & No \textsc{etl} / streaming ingestion for agent data &
Kinesis / Glue & \low \\
\bottomrule
\end{tabular}
\end{table}

\subsection{Category G: Agent Economic Primitives}
\label{subsec:gap-economy}

Agentverse includes a payment protocol and a basic marketplace, but lacks the
richer economic infrastructure that a mature agent economy requires.  The
search \textsc{api} returns agent listings with interaction counts and star
ratings, but there is no multi-dimensional trust score, no \textsc{sla}
enforcement mechanism, no automated pricing, and no revenue analytics for
agent developers.

We use the term \emph{agent economic primitives}---borrowing from mechanism
design \citep{myerson2007mechanism}---to describe the set of basic operations
that a functional agent economy requires: payment (exists), reputation
(rudimentary), lending, insurance, and price discovery (all absent).  Each has
a direct precedent in human financial systems and is tractable to implement on
a programmable blockchain.  Table~\ref{tab:gap-economy} enumerates the eight
economic primitive gaps.

\begin{table}[htbp]
\centering
\caption{Category G gaps: Agent Economic Primitives}
\label{tab:gap-economy}
\begin{tabular}{>{\raggedright\arraybackslash}p{3.2cm} >{\raggedright\arraybackslash}p{5.5cm} >{\raggedright\arraybackslash}p{2.9cm} l}
\toprule
\textbf{Gap} & \textbf{Description} & \textbf{Precedent} &
\textbf{Severity} \\
\midrule
No multi-dim.\ reputation & Single star rating only & eBay / Stripe Risk &
\high \\
No \textsc{sla} enforcement & No smart-contract-backed service guarantees & ---
& \high \\
No revenue analytics & No earnings and transaction history dashboard & Stripe
Dashboard & \high \\
No automated pricing & No dynamic pricing based on demand or capability & Spot
Instances & \medium \\
No resource lending & No borrowing of compute against future revenue & DeFi
lending & \low \\
No agent insurance & No risk pooling for agent failures or errors & Lloyd's
syndicates & \low \\
No dispute resolution & No arbitration for failed service delivery & eBay
Resolution & \medium \\
No KYA framework & No ``Know Your Agent'' compliance registry & KYC / AML &
\medium \\
\bottomrule
\end{tabular}
\end{table}

\subsection{Category H: Scale and Enterprise}
\label{subsec:gap-enterprise}

Agentverse currently supports a single execution instance per hosted agent.
There is no horizontal scaling, no multi-region deployment, no private
networking between agents, and no enterprise authentication integration.  The
platform has not published uptime \textsc{sla}s, compliance certifications, or
dedicated compute options.  These gaps are significant for enterprise adoption
in regulated industries---financial services, healthcare, legal---where
guaranteed uptime, data residency, and compliance certifications are
prerequisites.  Table~\ref{tab:gap-enterprise} details the eight scale and
enterprise gaps.

\begin{table}[htbp]
\centering
\caption{Category H gaps: Scale and Enterprise}
\label{tab:gap-enterprise}
\begin{tabular}{>{\raggedright\arraybackslash}p{3.2cm} >{\raggedright\arraybackslash}p{5.5cm} >{\raggedright\arraybackslash}p{2.9cm} l}
\toprule
\textbf{Gap} & \textbf{Description} & \textbf{Cloud Analogue} &
\textbf{Severity} \\
\midrule
No horizontal scaling & Single instance per agent; no auto-scaling &
\textsc{ecs} / Kubernetes & \high \\
No multi-region & Single deployment region only & \textsc{aws} multi-region &
\high \\
No private networking & All communication via public infrastructure & \textsc{vpc}
/ PrivateLink & \medium \\
No \textsc{sso} / enterprise \textsc{iam} & No \textsc{saml}, \textsc{oidc}, or
\textsc{scim} integration & Cognito / Okta & \high \\
No published uptime \textsc{sla} & No contractual uptime guarantee & \textsc{aws sla} &
\high \\
No compliance certifications & No SOC~2, ISO 27001, HIPAA & \textsc{aws}
Compliance & \high \\
No dedicated compute & No reserved or dedicated execution tier & Reserved
Instances & \medium \\
No data residency & No control over geographic data placement & \textsc{aws}
Regions & \medium \\
\bottomrule
\end{tabular}
\end{table}

\subsection{Summary}
\label{subsec:gap-summary}

Across the eight categories, we identify \textbf{62 distinct capability gaps}.
The concentration of High-severity gaps in Categories A (Memory), C (Security),
D (Networking), E (Development), F (Data Services), and H (Enterprise) reflects
the platform's current positioning as a capable prototype rather than a
production-grade infrastructure platform.  Hyperscaler platforms close most
Category F and H gaps by inheriting their parent cloud's managed service
catalogue, but lack Agentverse's advantages in Categories A (agent-specific
memory architecture), D (open communication protocol), and G (agent-native
economic primitives).  The architecture in Section~\ref{sec:architecture} is
designed to close the gaps while preserving these advantages.

\section{A Layered Architecture for the Agent Cloud: 2030}
\label{sec:architecture}

Drawing on the gap analysis of Section~\ref{sec:gaps} and the cloud computing
reference model of Section~\ref{subsec:cloud-reference}, we propose a
seven-layer reference architecture for a fully realised \textsc{aaas} (Agents
as a Service) cloud platform.  This stack is the \textsc{aaas} equivalent of
the \textsc{osi} model: a principled decomposition of every capability an
agent cloud must provide, from cryptographic identity at the substrate to
economic exchange at the application layer.  Each layer addresses one or more
gap categories from Section~\ref{sec:gaps}.  The layers are ordered from
infrastructure (Layer~0) to application (Layer~6), following the convention
of the \textsc{osi} model.  The complete stack is depicted in
Figure~\ref{fig:stack}.

\begin{figure}[htbp]
\centering
\newlength{\agstackrcw}%
\setlength{\agstackrcw}{\linewidth}%
\addtolength{\agstackrcw}{-4.1cm}%
\begin{tikzpicture}[
  font=\small,
  lbl/.style={anchor=west, text width=3.55cm, align=left,
              font=\small\bfseries, inner sep=0pt, outer sep=0pt},
  dsc/.style={anchor=west, text width=\agstackrcw, align=left,
              inner sep=0pt, outer sep=0pt}
]
\fill[blue!16!white]   (0,0.00cm) rectangle (\linewidth,1.15cm); 
\fill[green!13!white]  (0,1.15cm) rectangle (\linewidth,2.30cm); 
\fill[cyan!11!white]   (0,2.30cm) rectangle (\linewidth,3.45cm); 
\fill[teal!11!white]   (0,3.45cm) rectangle (\linewidth,4.60cm); 
\fill[yellow!18!white] (0,4.60cm) rectangle (\linewidth,5.75cm); 
\fill[orange!14!white] (0,5.75cm) rectangle (\linewidth,6.90cm); 
\fill[violet!13!white] (0,6.90cm) rectangle (\linewidth,8.05cm); 
\foreach \y in {1.15cm,2.30cm,3.45cm,4.60cm,5.75cm,6.90cm}{%
  \draw[black!30,line width=0.35pt] (0,\y) -- (\linewidth,\y);}
\draw[black!55,line width=0.6pt] (0,0cm) rectangle (\linewidth,8.05cm);
\draw[black!30,line width=0.35pt] (3.8cm,0cm) -- (3.8cm,8.05cm);
\node[lbl] at (0.15cm,0.575cm)  {Layer 0:\\Substrate};
\node[dsc] at (3.95cm,0.575cm)  {ASI Chain, DIDs, Verifiable Credentials, TEEs};
\node[lbl] at (0.15cm,1.725cm)  {Layer 1:\\Runtime};
\node[dsc] at (3.95cm,1.725cm)  {Wasm sandbox, capability-based security, auto-scaling};
\node[lbl] at (0.15cm,2.875cm)  {Layer 2:\\Memory};
\node[dsc] at (3.95cm,2.875cm)  {Episodic, Semantic, Procedural, Working, Shared (via MCP)};
\node[lbl] at (0.15cm,4.025cm)  {Layer 3:\\Communication};
\node[dsc] at (3.95cm,4.025cm)  {Multi-protocol gateway (Chat $+$ A2A $+$ MCP), event bus};
\node[lbl] at (0.15cm,5.175cm)  {Layer 4:\\Services};
\node[dsc] at (3.95cm,5.175cm)  {AgentDB, AgentStore, AgentCache, AgentMCP Hub};
\node[lbl] at (0.15cm,6.325cm)  {Layer 5:\\Observability};
\node[dsc] at (3.95cm,6.325cm)  {Tracing, cost dashboards, guardrail eval, sim env};
\node[lbl] at (0.15cm,7.475cm)  {Layer 6:\\Economy};
\node[dsc] at (3.95cm,7.475cm)  {Marketplace, reputation, KYA, economic primitives};
\end{tikzpicture}
\caption{The Seven-Layer Agent Cloud Stack (read bottom to top).  Layer~0
provides the decentralised trust substrate; Layers~1--6 add progressively
higher-level services, each closing specific gaps identified in
Section~\ref{sec:gaps}.}
\label{fig:stack}
\end{figure}

\subsection{Layer 0: Agent Substrate (Identity + Compute Foundation)}
\label{subsec:layer0}

\textbf{Identity substrate.} Every agent should have a W3C Decentralised
Identifier (\textsc{did}) \citep{garzon2025agentdid,w3c2022did} anchored to
the ASI Chain.  \textsc{did}s enable agents to prove identity across platform
boundaries and carry Verifiable Credentials (\textsc{vc}s) attesting to
capabilities, compliance certifications, and reputation.  The existing
\texttt{agent1q...} address format is conceptually compatible with \textsc{did}
anchoring; the ASI Chain provides the ledger needed for \textsc{did}
registration.

\textbf{Execution anchoring.} By 2030, agent execution records---start/stop
events, resource consumption, output hashes---should be anchored on the ASI
Chain, enabling permissionless verification of agent behaviour.  Trusted
Execution Environments (\textsc{tee}s) should be available for agents handling
sensitive data.

\subsection{Layer 1: Agent Runtime (Execution Environment)}
\label{subsec:layer1}

\textbf{Sandboxed execution.} Agent code should run in WebAssembly sandboxes
with capability-based security.  Each agent receives an explicit set of
capability tokens at instantiation---access to specific \textsc{mcp} servers,
specific secret names, specific network endpoints.  Anything not explicitly
granted is inaccessible.  This directly addresses Category~C gaps.

\textbf{Multi-language support.} The current Python-only restriction should be
lifted.  Wasm-compiled agents in TypeScript, Rust, and Go should be deployable
alongside Python agents, enabling a broader developer community.

\textbf{Auto-scaling.} Popular agents should scale horizontally---multiple
execution instances behind a load balancer---rather than being limited to a
single instance.  Auto-scaling requires that agent state be externalised
(handled by Layer~2), making the memory and runtime layers interdependent.

\subsection{Layer 2: Agent Memory (Persistence + Knowledge)}
\label{subsec:layer2}

The memory layer is the most consequential addition to the current platform.
We propose five services, each exposed as a per-agent authenticated
\textsc{mcp} server:

\textbf{Episodic Memory} (\texttt{memory://agent/\{address\}/episodic}):
Temporally indexed interaction histories---what happened, when, with whom, and
with what outcome.  Implemented as a time-series store with full-text and
embedding-based retrieval.

\textbf{Semantic Memory} (\texttt{memory://agent/\{address\}/semantic}): A
knowledge graph \citep{zepgraphiti2025} storing facts, concepts, and relationships
with temporal validity bounds.  Supports multi-hop graph traversal and hybrid
vector-graph retrieval.

\textbf{Procedural Memory} (\texttt{memory://agent/\{address\}/procedural}):
Learned workflows, tool usage patterns, and action sequences that have proven
effective.

\textbf{Working Memory} (\texttt{memory://agent/\{address\}/working}): A fast,
short-lived cache for the active context of ongoing sessions, automatically
evicted when the session ends.

\textbf{Shared Memory Spaces} (\texttt{memory://space/\{space\_id\}}):
Permissioned shared namespaces for multi-agent workflows.  Access governed by
Verifiable Credentials (Layer~0).

The critical architectural choice is that \textbf{all memory services are
exposed as \textsc{mcp} servers}.  This means any \textsc{mcp}-compatible
agent or client can access its memory using standard tool calls---no
Agentverse-specific \textsc{sdk} required.  Memory becomes a platform utility
with a universal interface.

\subsection{Layer 3: Agent Communication (Protocols + Networking)}
\label{subsec:layer3}

\textbf{Multi-protocol gateway.} A gateway that transparently bridges the
Agent Chat Protocol, A2A \citep{a2a2026}, and \textsc{mcp}
\citep{mcp2025spec}, enabling Agentverse agents to participate in A2A task
graphs and consume \textsc{mcp} services without leaving the platform.  This
addresses the most strategically significant Category~D gap.

\textbf{Event bus.} A pub/sub system---the agent-native equivalent of Amazon
\textsc{sns} or Google Pub/Sub---enabling broadcast coordination between agents
and reactive, event-driven architectures.

\textbf{Streaming support.} Messages should support Server-Sent Events and
WebSocket streaming for agents producing long-running outputs.

\textbf{Group coordination primitives.} Native support for consensus,
delegation, and quorum patterns, enabling autonomous multi-agent organisations.

\textbf{Human escalation protocol.} A standardised, auditable path for an
agent to request human oversight, satisfying EU AI Act requirements for
high-risk systems \citep{owasp2025agentic}.

\subsection{Layer 4: Agent Services (Managed Cloud Services)}
\label{subsec:layer4}

The managed service catalogue that closes Category~F gaps:

\begin{itemize}[leftmargin=*]
\item \textbf{AgentDB}: Hosted relational and document database, provisioned
per agent
\item \textbf{AgentStore}: Object storage for agent-generated assets
(\textsc{s3} equivalent)
\item \textbf{AgentCache}: Managed Redis-compatible in-memory cache
\item \textbf{AgentQueue}: Durable task queue for deferred work and backpressure
\item \textbf{AgentSecrets}: Per-agent secrets management with automatic
rotation
\item \textbf{AgentMCP Hub}: Marketplace of pre-built authenticated
\textsc{mcp} servers---web search, calendar, email, code execution, browser
control, financial data---subscribable with a single \textsc{api} call
\end{itemize}

\subsection{Layer 5: Agent Observability (Monitoring + Debugging)}
\label{subsec:layer5}

\begin{itemize}[leftmargin=*]
\item \textbf{Distributed tracing}: OpenTelemetry-native spans for every
\textsc{llm} call, tool invocation, memory access, and message, aggregatable
into execution trees
\item \textbf{Cost dashboards}: Per-agent visibility into token consumption,
compute seconds, storage, and payment flows
\item \textbf{Behaviour analytics}: Aggregate views of interaction patterns,
tool usage, and error clustering
\item \textbf{Guardrail evaluation}: Automated testing of agent outputs against
safety and quality criteria
\item \textbf{Simulation environment}: Sandboxed testing against synthetic
workloads and adversarial inputs
\end{itemize}

\subsection{Layer 6: Agent Economy (Marketplace + Economic Primitives)}
\label{subsec:layer6}

\textbf{Semantic Agent Discovery.} Natural-language capability search returning
agents ranked by capability match, reputation, price, and current
availability---the Almanac evolved from a phone book into a recommendation
engine.

\textbf{Multi-dimensional reputation.} Composable on-chain reputation scores
covering reliability, quality, safety, and speed---portable across any platform
supporting the \textsc{did}/\textsc{vc} trust model \citep{garzon2025agentdid}.

\textbf{SLA enforcement via smart contracts.} Service Level Agreements
automatically enforced by ASI Chain smart contracts, transforming the
marketplace from trust-based to trust-verified.

\textbf{Agent economic primitives} \citep{myerson2007mechanism}:
\begin{itemize}[leftmargin=*]
\item \emph{Resource lending}: Agents borrow compute capacity against future
revenue
\item \emph{Risk pooling}: Developers insure against agent failures via shared
reserves
\item \emph{Price discovery}: Dynamic rate negotiation based on demand, supply,
and reputation history
\end{itemize}

\textbf{Know Your Agent (KYA) compliance registry.}
Verifiable attestations of operator identity, intended use, safety
certifications, and compliance status---the agent equivalent of KYC
\citep{chainup2026kya}.

\section{Critical Evolution Paths}
\label{sec:paths}

We identify five specific transitions that Agentverse must undergo between its
current state and the 2030 architecture of Section~\ref{sec:architecture}.

\subsection{From ctx.storage to Agent Memory Cloud}
\label{subsec:path-memory}

\textbf{Current state.} \texttt{ctx.storage}---a flat key-value store with no
schema, querying, indexing, size visibility, or cross-agent sharing.

\textbf{2030 state.} A four-tier memory architecture (episodic, semantic,
procedural, working), plus shared memory spaces, all exposed as per-agent
authenticated \textsc{mcp} servers.

\textbf{The database analogy.} This transition is equivalent to the shift from
writing application data to local flat files to using a managed relational
database.  Before hosted databases, every web application implemented its own
persistence.  Hosted databases (\textsc{mysql}, then \textsc{rds}, then
DynamoDB) transformed what was possible at acceptable development cost.  Today
every Agentverse agent that needs memory must implement its own persistence
outside the platform; an Agent Memory Cloud would make sophisticated memory
available as a platform utility.

\textbf{The MCP interface.} Memory services exposed as \textsc{mcp} servers
means any \textsc{mcp}-compatible framework can give its agents persistent
memory by subscribing to the appropriate endpoint.  Memory becomes
infrastructure-level, not application-level.  Recent work by
\citet{zepgraphiti2025} on graph-based memory achieves 63.8\% accuracy on
LongMemEval versus 49.0\% for flat vector stores---a 30\% relative
improvement that would be immediately available to all Agentverse agents if
integrated at the memory layer.

\subsection{From Almanac to Semantic Agent DNS}
\label{subsec:path-dns}

\textbf{Current state.} The Almanac maps addresses to endpoints and protocols.
Search supports keyword and rudimentary semantic matching over the registered
agent population.

\textbf{2030 state.} A semantic, trust-weighted discovery system with real-time
capability negotiation: a \textsc{dns} plus search engine for the agent web.

\textbf{The analogy.} The early internet had WHOIS (address-to-information
lookup) before \textsc{dns} and Google.  Agents need the same evolution:
reliable address resolution for known agents, and semantic discovery for
finding the right agent for a novel task.  The 2030 Almanac should accept a
natural-language capability description and return agents ranked by match,
reputation, price, and availability---and should support real-time capability
negotiation between a requesting and a candidate agent.  The scale at which the
Almanac already operates---36,000+ agents as of Q1 2026
\citep{kantor2026stateofagents}---makes this capability urgent.

\subsection{From Chat Protocol to Agent Lingua Franca}
\label{subsec:path-protocol}

\textbf{Current state.} The Agent Chat Protocol is the native communication
standard.  A2A and \textsc{mcp} are supported only via external adapters, not
platform-integrated.

\textbf{2030 state.} A multi-protocol gateway bridging Chat Protocol, A2A
\citep{a2a2026}, and \textsc{mcp} \citep{mcp2025spec}, enabling Agentverse
agents to participate in any compliant multi-agent system.

\textbf{The \textsc{tcp/ip} parallel.} The internet protocol landscape of the
1980s was fragmented; \textsc{tcp/ip} won by becoming the translation layer.
By 2030, either one protocol dominates agent communication, or a gateway plays
the \textsc{tcp/ip} role.  The Agent Chat Protocol's typed, schema-validated
semantics are compatible with both A2A and \textsc{mcp}; the convergence may
be more about shared \emph{semantics}---a common ontology for expressing
capabilities, tasks, and results---than about wire format.  Agentverse is
uniquely positioned to build this gateway, as it has already committed to both
A2A adapter support and the most comprehensive \textsc{mcp} integration among
decentralised platforms.

\subsection{From Hosting to Agent Kubernetes}
\label{subsec:path-orchestration}

\textbf{Current state.} Each agent runs as a single instance with no health
monitoring, no scaling, no rolling deployments, and no multi-region distribution.

\textbf{2030 state.} Agents deployed as scalable units with desired replica
count, health probes, rolling update strategy, resource limits, and geographic
placement constraints.

\textbf{The dependency.} Horizontal scaling requires that agent state be
externalised.  An agent that stores state in-process (even via
\texttt{ctx.storage}, which is instance-local) breaks when replicated.  This
makes Layer~2 (Agent Memory) a prerequisite for Layer~1 (Agent Runtime)
scaling---a dependency that must be designed into the architecture from the
outset.

\subsection{From Token Payments to Agent Economic Primitives}
\label{subsec:path-economy}

\textbf{Current state.} The Payment Protocol supports Stripe, Skyfire
\textsc{usdc}, and \textsc{asi}/\textsc{fet} on-chain payments.  This is
operational---agents can charge for services and pay each other.

\textbf{2030 state.} Multi-dimensional reputation, \textsc{sla}-backed service
agreements, resource lending, insurance pools, and dynamic price discovery.

\textbf{The mechanism design framing.} We use \emph{agent economic primitives}
deliberately to avoid conflating this with token speculation.  Each primitive
solves a specific information or commitment problem in agent markets:
reputation reduces information asymmetry; \textsc{sla} contracts enable
credible commitment; insurance pools enable risk transfer; dynamic pricing
enables allocative efficiency.  These are mechanism design problems
\citep{myerson2007mechanism} with known theoretical solutions---the engineering
challenge is implementing them on a programmable blockchain at agent scale.
The emergence of KYA (Know Your Agent) frameworks \citep{chainup2026kya} and
dedicated agent marketplaces \citep{agenticmarket2026} suggests the ecosystem
is beginning to organise around exactly these needs.

\section{Discussion}
\label{sec:discussion}

\subsection{Competitive Positioning}
\label{subsec:competitive}

\begin{table}[htbp]
\centering
\caption{Comparative capabilities of agent platforms, 2026 vs.\ projected 2030.
Cell colour: \textcolor{green!60!black}{$\blacksquare$}~excellent,
             \textcolor{green!40!black}{$\blacksquare$}~good,
             \textcolor{orange!70!black}{$\blacksquare$}~moderate,
             \textcolor{orange!90!black}{$\blacksquare$}~limited,
             \textcolor{red!70!black}{$\blacksquare$}~minimal.
Last column (Agentverse 2030) separated by thicker rule.}
\label{tab:platform-comparison}
\begin{tikzpicture}[
  font=\scriptsize,
  lbl/.style={anchor=center, text width=2.65cm, align=left,
              font=\scriptsize\bfseries, inner sep=0pt, outer sep=0pt},
  hdr/.style={anchor=center, text width=1.73cm, align=center,
              font=\scriptsize\bfseries, inner sep=0pt, outer sep=0pt},
  dsc/.style={anchor=center, text width=1.73cm, align=center,
              inner sep=0pt, outer sep=0pt}
]

\fill[gray!6!white] (0,11.000cm) rectangle (2.800cm,12.100cm);
\fill[gray!18!white] (2.800cm,11.000cm) rectangle (12.200cm,12.100cm);
\node[lbl] at (1.400cm,11.550cm) {\textbf{Dimension}};
\node[hdr] at (3.740cm,11.550cm) {{Agentverse\\2026}};
\node[hdr] at (5.620cm,11.550cm) {{AWS Bedrock}};
\node[hdr] at (7.500cm,11.550cm) {{Azure AI\\Foundry}};
\node[hdr] at (9.380cm,11.550cm) {{Google\\Gemini Ent.}};
\node[hdr] at (11.260cm,11.550cm) {{Agentverse\\2030}};

\fill[gray!6!white] (0,9.900cm) rectangle (2.800cm,11.000cm);
\fill[orange!22!white] (2.800cm,9.900cm) rectangle (4.680cm,11.000cm);
\fill[green!15!white] (4.680cm,9.900cm) rectangle (6.560cm,11.000cm);
\fill[green!15!white] (6.560cm,9.900cm) rectangle (8.440cm,11.000cm);
\fill[green!15!white] (8.440cm,9.900cm) rectangle (10.320cm,11.000cm);
\fill[green!30!white] (10.320cm,9.900cm) rectangle (12.200cm,11.000cm);
\node[lbl] at (1.400cm,10.450cm) {Compute};
\node[dsc] at (3.740cm,10.450cm) {Hosted Python};
\node[dsc] at (5.620cm,10.450cm) {Serverless, managed};
\node[dsc] at (7.500cm,10.450cm) {Managed, scalable};
\node[dsc] at (9.380cm,10.450cm) {Managed, scalable};
\node[dsc] at (11.260cm,10.450cm) {Wasm, multi-lang};

\fill[gray!6!white] (0,8.800cm) rectangle (2.800cm,9.900cm);
\fill[green!30!white] (2.800cm,8.800cm) rectangle (4.680cm,9.900cm);
\fill[orange!22!white] (4.680cm,8.800cm) rectangle (6.560cm,9.900cm);
\fill[orange!22!white] (6.560cm,8.800cm) rectangle (8.440cm,9.900cm);
\fill[orange!22!white] (8.440cm,8.800cm) rectangle (10.320cm,9.900cm);
\fill[green!30!white] (10.320cm,8.800cm) rectangle (12.200cm,9.900cm);
\node[lbl] at (1.400cm,9.350cm) {Identity};
\node[dsc] at (3.740cm,9.350cm) {Crypto keys};
\node[dsc] at (5.620cm,9.350cm) {AWS IAM};
\node[dsc] at (7.500cm,9.350cm) {Entra Agent ID};
\node[dsc] at (9.380cm,9.350cm) {Google IAM};
\node[dsc] at (11.260cm,9.350cm) {DIDs + VCs};

\fill[gray!6!white] (0,7.700cm) rectangle (2.800cm,8.800cm);
\fill[red!22!white] (2.800cm,7.700cm) rectangle (4.680cm,8.800cm);
\fill[green!30!white] (4.680cm,7.700cm) rectangle (6.560cm,8.800cm);
\fill[green!30!white] (6.560cm,7.700cm) rectangle (8.440cm,8.800cm);
\fill[green!30!white] (8.440cm,7.700cm) rectangle (10.320cm,8.800cm);
\fill[green!30!white] (10.320cm,7.700cm) rectangle (12.200cm,8.800cm);
\node[lbl] at (1.400cm,8.250cm) {Memory};
\node[dsc] at (3.740cm,8.250cm) {ctx.storage};
\node[dsc] at (5.620cm,8.250cm) {S3 + DynamoDB};
\node[dsc] at (7.500cm,8.250cm) {Cosmos DB};
\node[dsc] at (9.380cm,8.250cm) {Firestore};
\node[dsc] at (11.260cm,8.250cm) {Memory Cloud (4-tier)};

\fill[gray!6!white] (0,6.600cm) rectangle (2.800cm,7.700cm);
\fill[yellow!25!white] (2.800cm,6.600cm) rectangle (4.680cm,7.700cm);
\fill[yellow!25!white] (4.680cm,6.600cm) rectangle (6.560cm,7.700cm);
\fill[green!15!white] (6.560cm,6.600cm) rectangle (8.440cm,7.700cm);
\fill[green!15!white] (8.440cm,6.600cm) rectangle (10.320cm,7.700cm);
\fill[green!30!white] (10.320cm,6.600cm) rectangle (12.200cm,7.700cm);
\node[lbl] at (1.400cm,7.150cm) {Communication};
\node[dsc] at (3.740cm,7.150cm) {Chat Protocol};
\node[dsc] at (5.620cm,7.150cm) {REST, Bedrock API};
\node[dsc] at (7.500cm,7.150cm) {REST, A2A};
\node[dsc] at (9.380cm,7.150cm) {A2A, REST};
\node[dsc] at (11.260cm,7.150cm) {Multi-protocol gateway};

\fill[gray!6!white] (0,5.500cm) rectangle (2.800cm,6.600cm);
\fill[yellow!25!white] (2.800cm,5.500cm) rectangle (4.680cm,6.600cm);
\fill[yellow!25!white] (4.680cm,5.500cm) rectangle (6.560cm,6.600cm);
\fill[yellow!25!white] (6.560cm,5.500cm) rectangle (8.440cm,6.600cm);
\fill[green!15!white] (8.440cm,5.500cm) rectangle (10.320cm,6.600cm);
\fill[green!30!white] (10.320cm,5.500cm) rectangle (12.200cm,6.600cm);
\node[lbl] at (1.400cm,6.050cm) {Discovery};
\node[dsc] at (3.740cm,6.050cm) {Almanac};
\node[dsc] at (5.620cm,6.050cm) {Bedrock Registry};
\node[dsc] at (7.500cm,6.050cm) {Copilot Studio};
\node[dsc] at (9.380cm,6.050cm) {Agentspace};
\node[dsc] at (11.260cm,6.050cm) {Semantic Agent DNS};

\fill[gray!6!white] (0,4.400cm) rectangle (2.800cm,5.500cm);
\fill[red!22!white] (2.800cm,4.400cm) rectangle (4.680cm,5.500cm);
\fill[green!30!white] (4.680cm,4.400cm) rectangle (6.560cm,5.500cm);
\fill[green!30!white] (6.560cm,4.400cm) rectangle (8.440cm,5.500cm);
\fill[green!30!white] (8.440cm,4.400cm) rectangle (10.320cm,5.500cm);
\fill[green!30!white] (10.320cm,4.400cm) rectangle (12.200cm,5.500cm);
\node[lbl] at (1.400cm,4.950cm) {Observability};
\node[dsc] at (3.740cm,4.950cm) {Log polling};
\node[dsc] at (5.620cm,4.950cm) {CloudWatch, X-Ray};
\node[dsc] at (7.500cm,4.950cm) {Azure Monitor};
\node[dsc] at (9.380cm,4.950cm) {Cloud Trace};
\node[dsc] at (11.260cm,4.950cm) {OTel + cost dashboards};

\fill[gray!6!white] (0,3.300cm) rectangle (2.800cm,4.400cm);
\fill[orange!22!white] (2.800cm,3.300cm) rectangle (4.680cm,4.400cm);
\fill[green!30!white] (4.680cm,3.300cm) rectangle (6.560cm,4.400cm);
\fill[green!30!white] (6.560cm,3.300cm) rectangle (8.440cm,4.400cm);
\fill[green!30!white] (8.440cm,3.300cm) rectangle (10.320cm,4.400cm);
\fill[green!30!white] (10.320cm,3.300cm) rectangle (12.200cm,4.400cm);
\node[lbl] at (1.400cm,3.850cm) {Security};
\node[dsc] at (3.740cm,3.850cm) {Account-level};
\node[dsc] at (5.620cm,3.850cm) {IAM, VPC, KMS};
\node[dsc] at (7.500cm,3.850cm) {Entra + Azure Sec.};
\node[dsc] at (9.380cm,3.850cm) {Google IAM + VPC};
\node[dsc] at (11.260cm,3.850cm) {Capability-based, Wasm};

\fill[gray!6!white] (0,2.200cm) rectangle (2.800cm,3.300cm);
\fill[green!30!white] (2.800cm,2.200cm) rectangle (4.680cm,3.300cm);
\fill[red!22!white] (4.680cm,2.200cm) rectangle (6.560cm,3.300cm);
\fill[red!22!white] (6.560cm,2.200cm) rectangle (8.440cm,3.300cm);
\fill[red!22!white] (8.440cm,2.200cm) rectangle (10.320cm,3.300cm);
\fill[green!30!white] (10.320cm,2.200cm) rectangle (12.200cm,3.300cm);
\node[lbl] at (1.400cm,2.750cm) {Payments};
\node[dsc] at (3.740cm,2.750cm) {ASI + Stripe + USDC};
\node[dsc] at (5.620cm,2.750cm) {AWS Billing};
\node[dsc] at (7.500cm,2.750cm) {Azure Credits};
\node[dsc] at (9.380cm,2.750cm) {GCP Billing};
\node[dsc] at (11.260cm,2.750cm) {Econ. primitives};

\fill[gray!6!white] (0,1.100cm) rectangle (2.800cm,2.200cm);
\fill[green!30!white] (2.800cm,1.100cm) rectangle (4.680cm,2.200cm);
\fill[red!22!white] (4.680cm,1.100cm) rectangle (6.560cm,2.200cm);
\fill[red!22!white] (6.560cm,1.100cm) rectangle (8.440cm,2.200cm);
\fill[red!22!white] (8.440cm,1.100cm) rectangle (10.320cm,2.200cm);
\fill[green!30!white] (10.320cm,1.100cm) rectangle (12.200cm,2.200cm);
\node[lbl] at (1.400cm,1.650cm) {Decentralisation};
\node[dsc] at (3.740cm,1.650cm) {$\bigstar\bigstar\bigstar\bigstar\bigstar$};
\node[dsc] at (5.620cm,1.650cm) {$\bigstar\bigcirc\bigcirc\bigcirc\bigcirc$};
\node[dsc] at (7.500cm,1.650cm) {$\bigstar\bigcirc\bigcirc\bigcirc\bigcirc$};
\node[dsc] at (9.380cm,1.650cm) {$\bigstar\bigcirc\bigcirc\bigcirc\bigcirc$};
\node[dsc] at (11.260cm,1.650cm) {$\bigstar\bigstar\bigstar\bigstar\bigstar$};

\fill[gray!6!white] (0,0.000cm) rectangle (2.800cm,1.100cm);
\fill[orange!22!white] (2.800cm,0.000cm) rectangle (4.680cm,1.100cm);
\fill[green!30!white] (4.680cm,0.000cm) rectangle (6.560cm,1.100cm);
\fill[green!30!white] (6.560cm,0.000cm) rectangle (8.440cm,1.100cm);
\fill[green!15!white] (8.440cm,0.000cm) rectangle (10.320cm,1.100cm);
\fill[green!15!white] (10.320cm,0.000cm) rectangle (12.200cm,1.100cm);
\node[lbl] at (1.400cm,0.550cm) {Enterprise};
\node[dsc] at (3.740cm,0.550cm) {$\bigstar\bigstar\bigcirc\bigcirc\bigcirc$};
\node[dsc] at (5.620cm,0.550cm) {$\bigstar\bigstar\bigstar\bigstar\bigstar$};
\node[dsc] at (7.500cm,0.550cm) {$\bigstar\bigstar\bigstar\bigstar\bigstar$};
\node[dsc] at (9.380cm,0.550cm) {$\bigstar\bigstar\bigstar\bigstar\bigcirc$};
\node[dsc] at (11.260cm,0.550cm) {$\bigstar\bigstar\bigstar\bigstar\bigcirc$};

\draw[black!55, line width=0.65pt] (0,0) rectangle (12.200cm,12.100cm);
\draw[black!55, line width=0.55pt] (0,11.000cm) -- (12.200cm,11.000cm);
\draw[black!22, line width=0.3pt] (0,9.900cm) -- (12.200cm,9.900cm);
\draw[black!22, line width=0.3pt] (0,8.800cm) -- (12.200cm,8.800cm);
\draw[black!22, line width=0.3pt] (0,7.700cm) -- (12.200cm,7.700cm);
\draw[black!22, line width=0.3pt] (0,6.600cm) -- (12.200cm,6.600cm);
\draw[black!22, line width=0.3pt] (0,5.500cm) -- (12.200cm,5.500cm);
\draw[black!22, line width=0.3pt] (0,4.400cm) -- (12.200cm,4.400cm);
\draw[black!22, line width=0.3pt] (0,3.300cm) -- (12.200cm,3.300cm);
\draw[black!22, line width=0.3pt] (0,2.200cm) -- (12.200cm,2.200cm);
\draw[black!22, line width=0.3pt] (0,1.100cm) -- (12.200cm,1.100cm);
\draw[black!55, line width=0.5pt] (2.800cm,0) -- (2.800cm,12.100cm);
\draw[black!22, line width=0.3pt] (4.680cm,0) -- (4.680cm,12.100cm);
\draw[black!22, line width=0.3pt] (6.560cm,0) -- (6.560cm,12.100cm);
\draw[black!22, line width=0.3pt] (8.440cm,0) -- (8.440cm,12.100cm);
\draw[black!22, line width=0.3pt] (10.320cm,0) -- (10.320cm,12.100cm);
\draw[black!50, line width=0.5pt] (10.320cm,0) -- (10.320cm,12.100cm);

\end{tikzpicture}
\end{table}

Table~\ref{tab:platform-comparison} reveals a structural complementarity:
hyperscaler platforms are strong on operational depth and weak on agent-native
properties; Agentverse is the inverse.  The competitive risk for Agentverse is
not that hyperscalers build better agent hosting but that they integrate A2A so
deeply into enterprise deployments that developers never feel the need to
register on an open platform.  The multi-protocol gateway (Section~\ref{subsec:path-protocol})
is the mitigation: if an Agentverse agent can be discovered and called from any
A2A client, the platform's openness becomes a feature rather than an island.

It is worth noting the pace of hyperscaler investment: Amazon Bedrock AgentCore
reached general availability in October 2025; Microsoft launched Entra Agent
ID into public preview at Ignite 2025 and announced expanded Agent 365
capabilities at RSAC 2026; Google consolidated Vertex AI and Agentspace into a
unified Gemini Enterprise Agent Platform at Cloud Next 2026.  The window in
which Agentverse can close the managed-service gap while retaining its
decentralised identity advantage is real but not indefinite.

\subsection{Open Research Problems}
\label{subsec:open-problems}

\textbf{Agent identity portability.} Can a \textsc{did} anchored on the ASI
Chain be resolved and verified on AWS without that platform running a full ASI
Chain node?  This is a cross-chain identity resolution problem analogous to
federated identity in Web2.

\textbf{Memory privacy in multi-agent systems.} If multiple agents share a
memory space, what privacy guarantees can be made about information contributed
by one agent from access by another?  Federated learning approaches exist but
their application to fine-grained agent memory is not well studied.

\textbf{Agent coordination theory.} The group coordination primitives we
propose (consensus, delegation, quorum) require a theory of multi-agent
coordination that goes beyond request-response.  Game-theoretic models provide
foundations; the specific dynamics of heterogeneous, incentive-driven agent
networks require further theoretical development.

\textbf{Verification of agent behaviour.} How do we provide formal or
probabilistic guarantees about what a deployed agent will and will not do, when
the agent includes non-deterministic \textsc{llm} components?  Testable
behavioural contracts and simulation-based evaluation (Layer~5) offer partial
solutions.

\textbf{Economic stability of agent markets.} The economic primitives we
propose must be incentive-compatible.  Mechanism design provides the
theoretical tools \citep{myerson2007mechanism}; their application to autonomous
agent markets with heterogeneous participants is an open problem.

\subsection{Regulatory Considerations}
\label{subsec:regulatory}

The EU AI Act (Regulation (EU)~2024/1689; high-risk provisions in force from
August~2, 2026) classifies
\textsc{ai} systems by risk level.  Autonomous agents taking consequential
actions---financial transactions, medical decisions, legal submissions---are
likely to be classified as high-risk, triggering requirements for human
oversight, auditability, and conformity assessment.  The architecture we
propose anticipates these requirements: the audit trail (Layer~5), human
escalation protocol (Layer~3), capability-based permissions (Layer~1), and
KYA registry (Layer~6) collectively provide the technical foundations for
compliance.  We argue for a mandatory baseline---every hosted agent should have
an immutable action log and a human escalation path---with optional higher
compliance tiers for regulated industries.

\subsection{Limitations}
\label{subsec:limitations}

Our analysis has four principal limitations.  First, it is grounded in a
single platform at a specific time; some gaps may not generalise to platforms
with different architectural choices.  Second, the 2030 projections are
conditional on current trajectories and could be materially altered by
technical breakthroughs or regulatory intervention.  Third, the economic
primitives in Sections~\ref{subsec:layer6} and~\ref{subsec:path-economy} are
proposed without formal analysis of their incentive properties; rigorous
mechanism design treatment of each would require dedicated theoretical work.
Fourth, while we supplement the endpoint audit with latency measurements
and search-coverage counts (Section~\ref{subsec:benchmarks}), the
performance data reported here are indicative rather than definitive:
we sample $n{=}10$ requests per endpoint from a single cloud region, which
is insufficient to characterise throughput under load, geographic latency
variability, or storage throughput.  The search-corpus counts are snapshot
figures that will shift as the registry grows.  More rigorous, multi-region
load tests would materially strengthen the quantitative contribution.

\section{Conclusion}
\label{sec:conclusion}

We have presented a systematic analysis of the Agentverse platform as a case
study in agent-native cloud infrastructure.  Our empirical audit of 204 \textsc{api} endpoints (as documented in Q1~2026)
reveals 62 capability gaps across eight
categories---from agent memory and observability to security, communication,
data services, economic primitives, and enterprise-grade scaling.  These gaps
are not failures of vision but the natural condition of pioneering
infrastructure: foundations take time, and Agentverse has laid important ones.

Our seven-layer Agent Cloud Stack provides a reference architecture for closing
these gaps by 2030.  Each layer is grounded in specific identified gaps and
informed by the evolution of general-purpose cloud computing as a reference
model.  The architecture preserves and extends the agent-native
properties---cryptographic identity, open protocols, decentralised
substrate---that distinguish Agentverse from its hyperscaler competitors, while
adding the operational depth (memory, observability, security, enterprise
services) that production deployments require.

The five critical evolution paths provide a concrete roadmap: from
\texttt{ctx.storage} to Agent Memory Cloud; from keyword Almanac to Semantic
Agent DNS; from single-protocol messaging to multi-protocol lingua franca; from
single-instance hosting to agent-native orchestration; and from token payments
to a full suite of agent economic primitives.  Each path is a discrete
engineering investment with a clear value proposition.

The broader significance lies in what this infrastructure enables.
\citet{hussein2025forecast} project more than $100\times$ growth in the global
agent population between 2026 and 2036.  The infrastructure that governs how
those agents interact---how they discover each other, remember their history,
communicate across platforms, verify each other's identity, and transact
economically---will shape outcomes in commerce, science, and governance far
beyond the scope of any individual agent application.  Agentverse, uniquely
among current platforms, is architecturally positioned to provide that
infrastructure in an open, decentralised, and composable form.  The seven-layer
stack we propose constitutes the full \textsc{aaas} (Agents as a Service)
model: the principled service abstraction for the agentic era, as foundational
to the agent cloud as \textsc{iaas} was to the compute cloud.  The distance
between where it is today and where it needs to be by 2030 is large but
well-defined.  This paper is a map of that distance.

\vspace{1em}
\noindent\textbf{Acknowledgements.}
The authors thank the Fetch.ai Foundation team for their work building the
Agentverse platform and for public discussions on the agentic web vision that
informed this analysis.  This work was carried out independently; any errors
or omissions are the authors' own.

\medskip
\noindent\textbf{AI Assistance Disclosure.}
This paper was researched and written with the assistance of
\textsc{ai} tools, including the Taurus multi-agent orchestration platform
(powered by Anthropic Claude) for literature synthesis, API exploration,
empirical measurement, and manuscript preparation.  The scientific framing,
gap taxonomy, architectural proposals, and all conclusions were designed,
evaluated, and verified by the authors.  The authors take full responsibility
for all claims, citations, and interpretations.


\bibliography{agentverse-paper}

\end{document}